\documentclass[apjl,twocolumn]{openjournal}

\makeatletter
\long\def\frontmatter@abstractheading{%
  \vspace*{-0.01in}%
  \centerline{\itshape\footnotesize\@submitted}
  \vspace*{0.11in}
  \begingroup
    \centering
    \abstractname
    \vskip 1mm
    \par
  \endgroup
  \everypar{\rightskip=0.5in\leftskip=\rightskip}\par
}
\makeatother

\shorttitle{Impact of Primordial Black Holes on Exoplanet Systems}
\shortauthors{Brown, He \& Unwin}

\journalinfo{The Open Journal of Astrophysics}
\submitted{Authors listed alphabetically; all authors contributed equally.}
\usepackage{amsmath,amsfonts}
\usepackage{hyperref}
\hypersetup{colorlinks=true, citecolor=blue, linkcolor=blue, urlcolor=blue}
\usepackage{xcolor}

\newcommand{\beq}{\begin{equation}\begin{aligned}}
\newcommand{\eeq}{\end{aligned}\end{equation}}

\makeatletter
\def\frontmatter@title@above{%
  \vspace*{0pt}%
  \footnotesize
  {\footnotesize\textsc{\@journalinfo}}\par
  {\scriptsize Preprint typeset using \LaTeX\ style openjournal v.\ \openjournaltemplate@ver}\par
  \vspace*{0.50in}
}%
\def\frontmatter@title@below{\vspace*{0.18in}}%
\makeatother

\begin{document}
\vspace*{-2mm}
\title{The Potential Impact of Primordial Black Holes on Exoplanet Systems}

\author{Garett Brown}
\affiliation{Department of Physical and Environmental Sciences, University of Toronto at Scarborough, Toronto, Ontario M1C 1A4, Canada}
\affiliation{Department of Physics, University of Toronto, Toronto, ON M5S 1A7, Canada}

\author{Linda He}
\affiliation{Harvard University, Cambridge, MA 02138, USA}

\author{James Unwin}

\affiliation{Department of Physics, University of Illinois Chicago, Chicago, IL 60607, USA}
\date{\today}

\vspace*{3mm}


\begin{abstract} 
The orbits of planetary systems can be deformed from their initial configurations due to close encounters with larger astrophysical bodies. Typical candidates for close encounters are stars and binaries. We explore the prospect that if there is a sizeable population of primordial black holes (PBH) in our galaxy, then these may also impact the orbits of exoplanets. Specifically, in a simplified setting, we study numerically how many planetary systems might have a close encounter with a PBH, and analyze the potential changes to the orbital parameters of systems that undergo PBH flybys. 
\end{abstract}


\section{Introduction}

In many cases, the interaction of a planet-star system with its environment can be reduced to a gravitational three-body problem. One such situation is when the third body makes a passing flyby, which perturbs the planetary system. Such flybys exchange energy with the planet-star system and may perturb the orbit of the planet. Other scenarios include collision between the transient flyby with the planet or the dissociation of the planet-star system; for a review, see e.g.~\cite{Cuello}. There are a variety of free-floating bodies which can potentially intrude on a stellar system, including free-floating planets, planetesimals, main sequence stars, brown dwarf stars, astrophysical black holes and binaries, as well as hypothetical populations of primordial black holes (PBH) or dark matter microhalos. 

There have been many studies of how the eccentricity and semi-major axis distributions for exoplanetary systems may be impacted by encounters with stars. Good starting points in the existing literature on close encounters between stellar systems and intruding bodies include \cite{1975AJ.....80..809H} and \cite{2015MNRAS.448..344L}. Moreover, \cite{2009ApJ...697..458S} explored analytical estimates of the changes in planetary orbital eccentricity and semi-major axis due to adiabatic and impulsive encounters with stellar bodies. 
Our aim is to extend the study of close encounters to consider more exotic hypothetical astrophysical bodies, focusing on PBH. One reason that populations of exotic objects are interesting is due to the fact that their typical mass scale, spatial distribution, and velocity dispersions may be very different to conventional objects such as stars and rouge planets. 
PBH are black holes which form not through stellar collapse, but rather from extreme overdensities in the early universe \citep{Novikov,Hawking:1971ei}, as such PBH can have masses well below one Solar Mass ($M_\odot$). 

While we will phrase our study in terms of PBH, our conclusions should be robust for other compact massive objects since the results are entirely set via their gravitational influence. Examples of other comparable hypothetical bodies include: ultracompact dark matter microhalos, axion minihalos, and dark matter stars \citep{Hogan:1988mp,Berezinsky:2013fxa,Freese:2015mta}. 
In the case of PBHs, because these objects form prior to galaxy formation, it is a reasonable expectation for the PBHs to have the same velocity dispersion as dark matter. Notably, since the mean of the dark matter velocity dispersion is $220$ km/s \citep{Navarro:1996gj}, the case of PBH can be highly distinct to the stellar case (with mean of $\sim40$ km/s). High velocity flybys, as would typically be the case for PBHs, are categorized as `impulsive encounters' \citep{2009ApJ...697..458S}.

\begin{figure}
    \centering
    \includegraphics[width=0.44\textwidth]{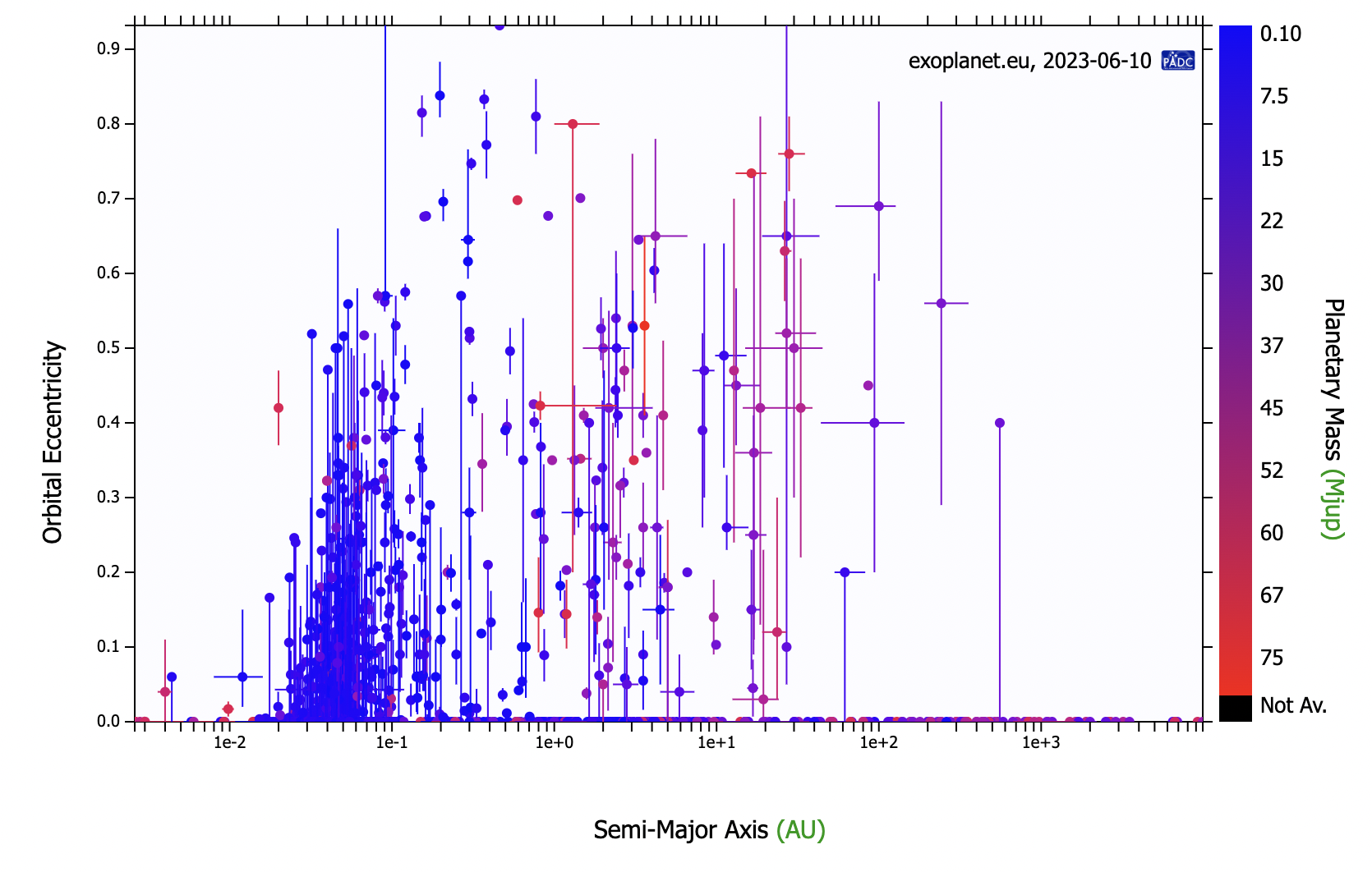}
\vspace{-3mm}
    \caption{Plot made through ``exoplanets.eu'' with the selection ``mass:mjup $>$ 0.1 AND "confirmed" in planet$\_$status''.}
    \label{fig:exo}
\end{figure}

In principle, if one could achieve high precision measurements and modelling of the distributions of exoplanet orbital parameters, then this could be used to infer or constrain the abundance of PBH. In practice, however, the large uncertainties relating to both measurements and the complex dynamical history of planetary systems present an obstacle to extracting constraints.  
On the positive side, the catalogues of known exoplanets have expanded rapidly over the last decade \citep{2011PASP..123..412W,2018ApJS..235...38T,Winn:2014xna}.  Though the first exoplanet was discovered in 1995 around a main-sequence star  \citep{Mayor1995}, there are now well over 5000 confirmed exoplanets in thousands of systems. These planets span a wide range of (dynamical) masses, semi-major axes,  eccentricities, and other orbital elements. Figure \ref{fig:exo} shows the range of semi-major axes and eccentricities of all currently confirmed exoplanets with masses in excess of 10\% of Jupiter. This was created using the tools available through \cite{exoplanets}.


Due to observational bias and dynamical complexity, there is not a complete understanding of the distributions of the orbital elements of exoplanets \citep{Burke2015, Christiansen2016}.  With these biases, in the current population of known exoplanets, Jupiter-like planets are the most common, in part because they are the easiest to detect. The most successful detection method, the transit method, is heavily biased towards short-period planets. There is also still much to understand regarding connections between proto-planetary disks and features of exoplanet systems \citep{Mulders2020, Emsenhuber2021}. 

Setting aside the challenges in observations and formation modeling, it is interesting to consider how late-time planetary orbits may be shaped due to interactions between planetary systems and transient close encounters with massive exotic astrophysical bodies that intrude into the parent star's radius of influence. Here we present a first analysis of this interesting prospect. 

This paper is organized as follows: Section \ref{sec:REBOUND} discusses the role of numerical simulations in studying close encounters. In Section \ref{sec3} we discuss the impact of stellar close encounters on exoplanet systems.  In Section \ref{sec4.0} we consider a galactic scale simulation to estimate the expected number of exoplanets over the entire galaxy that could experience a significant close encounter with a PBH. Section \ref{sec7}  describes our approach to modeling close encounters with PBH, and we simulate various scenarios for PBH flybys.  We also outline the idea that the statistical distribution of orbital parameters of exoplanets could be shaped by a galactic population of PBHs, and we present a simplified analysis for the case with $M_{\rm PBH}=0.1M_\odot$. Section \ref{limit}  discusses the limitations of our studies, and we provide concluding remarks in Section~\ref{conc}.


\vspace{-3mm}
\section{Simulating Close Encounters}
\label{sec:REBOUND} 

We begin by discussing how one can model close encounters and their impacts on exoplanet systems. Specifically, our intention is to study the outcomes of encounters between a flyby object and a simple planet-star system. The planet-star system consists of a central star with a single orbiting planet with semi-major axis $a$, and eccentricity $e$. We shall focus on the changes to the eccentricity $\Delta e$ or semi-major axis $\Delta a$. These can be related to the relative change in binding energy $\Delta E/E$ as follows (see e.g.~\citep{2015MNRAS.448..344L})
\beq
    \frac{\Delta E}{E} &= - \frac{\Delta a}{a}~,
    \label{eq1a}
    \eeq
    and to the relative change in the angular momentum
    \beq
    \frac{\Delta J}{J}  &= -  \frac{1}{2} \frac{\Delta a}{a} - \frac{e \Delta e }{1-e^2}~.
\label{eq1b}\eeq
Notably, parameters that strongly influence the perturbation strength of encounters are the distance of the closest approach, $r_{p}$, and the initial velocity of the intruder body relative to the planet-star system $v_0$. If $|v_0|$ is much larger than the orbital speed of the planet, then the encounter is said to be \textit{impulsive}, otherwise, the encounter is called   \textit{adiabatic} \citep{2009ApJ...697..458S}.  In the case of PBH, unlike typical stellar flybys, the relative velocity of PBHs is expected to be significantly larger than the orbital speed of the planet. Thus, we will be in the regime of impulsive encounters.

The parameters that describe the planet-star system are the parent star's mass $M_\star$, and the planet's mass $M_p$, semi-major axis $a$, and eccentricity $e$.
We introduce the passing flyby on a hyperbolic trajectory with eccentricity $\epsilon$ and impact parameter $b$, such that the intruder's closest approach to the star $r_{p}$ is given by
\begin{equation}\label{rp}
    r_{p} = b \sqrt{\frac{\epsilon-1}{\epsilon+1}} \approx b~.
\end{equation}
 The parameters that describe the flyby are the intruder's mass $M_I$, the flyby's initial velocity at infinity relative to the planet-star system $v_0$, and the impact parameter $b$. 
 
Analytic estimates have been studied for impulsive encounters in \cite{2009ApJ...697..458S}. This study found that impulsive encounters give the relative changes for the binding energy and angular momentum as stated in eqns.~(\ref{eq1a})~\&~(\ref{eq1b}).
The time scale for the encounter is $\tau \sim r_p / v$, where $v$ is the velocity of the flyby at the pericentral distance $r_p$. 
The change of velocity of the orbiting planet due to the flyby is 
\begin{equation}\label{dif}
    \delta v \approx \tau |\delta \ddot x_p - \delta \ddot x_\star | = \frac{4GM_I a}{v r_p^2}~,
\end{equation}
where $\delta \ddot x_p$ is the perturbation to the acceleration of the planet due to the intruder and $\delta\ddot x_\star$ is the acceleration of the parent star by the intruder.

Here, rather, we adopt a numerical approach to studying such impulsive close encounters.
To carry out simulations of close encounters between passing flybys with planet-star systems, we use the \texttt{REBOUND} \citep{Hanno:2012} software package, an N-body integrator that integrates the motion of particles under the influence of gravity.
Specifically, we adapt the package \texttt{REBOUND} to simulate impulsive close encounters. To our knowledge, this is the first such statistical study of impulsive close encounters.\footnote{While this work in preprint, \cite{Tran:2023jci} appeared, which explored complementary ideas, focused on our Solar System.}

We implement  \texttt{REBOUND} using a hybrid integration scheme\footnote{The code, titled ``{\tt airball}'', is available at \citep{airball}.} for switching from \texttt{WHCKL} \citep{ReinTamayoBrown2019} to \texttt{IAS15} \citep{IAS15} and back to \texttt{WHCKL} if the flyby object passes within $30a$ (thirty times the semi-major axis of the planet). If $r_p > 30a$ then \texttt{REBOUND}  uses \texttt{WHCKL} for the entire integration.
This hybrid scheme allows for fast integrations and high-resolution 3-body interactions because it takes advantage of the predominantly Keplerian motion when the bodies are weakly interacting, but uses the highly accurate, adaptive timestepping \texttt{IAS15} to resolve close encounters.

Our code uses a sufficiently small fixed timestep, between 1\% and 5\% of the innermost orbital period when integrating with \texttt{WHFast} \citep{2015MNRAS.452..376R}, so that when switching to \texttt{IAS15} the change in energy to the system incurred by switching integrators is negligible compared the change in energy due to the flyby.
Using the additional symplectic correctors and kernel method of \texttt{WHCKL} further reduces the incurred change in energy.
Switching back from \texttt{IAS15} to \texttt{WHFast} is also done at the same distance away from the planet-star system. 

The advantage of switching is being able to use the adaptive timestepping of \texttt{IAS15} to untangle the close encounters and strong interactions that break the assumptions of the \texttt{WHFast} integrator.
Our choice of switching integrators at $30a$ means that the \texttt{WHFast} assumptions always remain valid and in a regime where changes in energy of the planet due to switching integrators would be significantly less than the change in energy from the flyby object.

\begin{figure}
    \centering
    \includegraphics[width=0.4\textwidth]{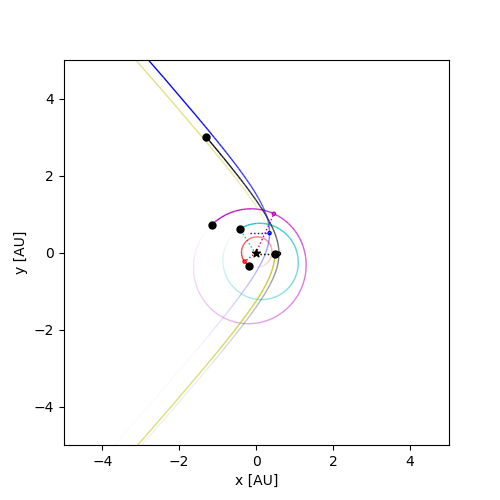}
\vspace{-2mm}
    \caption{For illustrative purposes only we show here the case of four planets on circular orbits around a central star ($\star$), whereas in our main simulations we only consider single planets orbiting stars. The flyby of the intruding body lies on hyperbolic trajectories, as shown as brown, black and yellow curves.  Coloured dots show the point of closest approach between certain pairs.
    \label{fig:sim}}
\end{figure}

We initialize \texttt{REBOUND} for a flyby object on a hyperbolic trajectory given the flyby's mass, velocity at infinity, and impact parameter.
Adding an object to a \texttt{REBOUND} simulation with a fully determined initial orbit requires the object's semi-major axis, eccentricity, inclination $i$, longitude of the ascending node $\Omega$, argument of perihelion $\omega$, and true anomaly $\nu$. Figure \ref{fig:sim}  illustrates the trajectories traced out under our simulation for a single run.

\section{Stellar Close Encounters}
\label{sec3}

The vast majority of stars are thought to form within clusters, and within these environments, stellar close encounters are relatively common \citep{Lada}. These interactions can disturb planetary systems, inducing gravitational perturbations that can result in significant changes in orbital eccentricities and inclinations of their constituent planets. Moreover, these encounters can also lead to the ejection of planets from their parent systems or their capture into new orbits. The gravitational influence of passing stars may also induce secular interactions, driving long-term changes in orbital parameters. Thus, stellar close encounters in birthing clusters are thought to play an important role in shaping the structure and dynamics of planetary systems. (See e.g.~\cite{Cuello} for a review.) Additionally, stellar systems will undergo close encounters with other stars, and binary star pairs in the galactic field \citep{2015MNRAS.448..344L}. 

The rate of stellar close encounters for a given star will depend on the local stellar density $\rho_*$. For young stellar clusters  in the Milky Way $\rho_*$ ranges from 0.01 to $10^5~M_\odot~{\rm pc}^{-3}$. The typical radial size of these clusters varies from 0.1 pc to $\mathcal{O}(10)$ pc \citep{Pfalzner}. 
Several groups have looked at estimating the likelihood of a star experiencing close encounters, e.g.~\citep{Munoz,Winter}.  

Let us first make our own simple scaling estimate via a Poisson distribution to provide an intuitive understanding of the likelihood of a single flyby event (or, indeed, $N$ events) within some time period. 
One can estimate the number of encounters in terms of an encounter cross section $\Gamma\simeq n\sigma v$. 
We use that the number density of stars in the birth cluster is roughly $n_\star\sim10^4 {\rm pc}^{-3}$ \citep{Kuhn}. 
We assume a geometric cross section $\sigma=\pi R^2$ of the radius at which stellar close encounters start to have an impact on exoplanet orbits, 
taking $R=1000$ AU. The duration a star spends in the cluster is of order $t_c\sim r_{\rm bc}/v_\star$ where we take the velocity of stars to be $\sim1$~km/s,
and we use that the typical size of a birthing cluster is $r_{\rm bc}\sim1$ pc. It follows that the average number of events $N$ that occur in the cluster is of order
\beq
N\sim\Gamma_\star t_c\simeq n_\star \sigma r_{\rm bc} \sim 0.7~.
\eeq
Moreover, the probability $P_N$ of exactly $N$ flybys to occur in time $t$ can be estimated from the Poisson distribution
\beq
P_N=\frac{1}{N!}(\Gamma t)^N\exp(-\Gamma t)~.
\label{PK}
\eeq
Taking $t=t_c$,  in Table \ref{table:tab1} we give the probability that a star experiences 1, 2, or 3 stellar flybys while in the birthing cluster. While this is a simplified analysis of a complex system, it gives some reasonable intuition.

\begin{table}
\centering 
\begin{tabular}{|c | c|} 
\hline
Number of Stellar Flybys ($k$) & Probability ($P_{\star k}$) \\ 
\hline 
1 &  0.35 \\[1pt]
2 & 0.12\\[1pt]
3 & 0.03 \\
\hline 
\end{tabular}
\caption{\normalfont{Probability $P_{\star k}$ of $k$ stellar close encounters for a given star in a cluster, assuming these events are Poisson distributed and with the parameters stated in the text.}}\label{table:tab1} 
\end{table}

Comparing to the literature: \cite{Winter}  considered clusters of uniform density of age 3 million years, composed of 1 $M_\odot$ stars with velocity dispersion $\sigma_*\sim 4 {\rm km/s}$ and $\rho_*\sim500$ pc${}^{-3}$ and estimated there was 70\% probability a star would pass within 1000 AU of another star. \cite{Pfalzner} used N-body simulations and found that more than half of stars undergo sub-1000 AU stellar encounters. 

These stellar close encounters have been shown to impact the orbital parameters of planetary systems. The semi-major axes of planets can be altered by 15\% - 40\%, eccentricities increase up to values of $e \sim 0.4$, and orbits may become inclined up to $\sim10^\circ$ \citep{Cuello,2014A&A...564A..28P,2019AJ....158...94B,2020ApJ...901...92M}. Further, planets or material in proto-planetary disks may be ejected \citep{Hills,2013ApJ...769..150C,2022MNRAS.514..920D}, leading to disk truncation. 

Prior to observations, planet-formation theory favoured coplanar and circular orbits, since even if a body acquired a moderate eccentricity or inclination, subsequent interactions with the disk were expected to erase these \citep{Cresswell,Xiang}.  Observations, however,  did not conform to these initial expectations \citep{2011PASP..123..412W,2018ApJS..235...38T}, this is thought to be due to the rich dynamics which occur after planets form such as planet-planet scattering and close encounters with other massive bodies \citep{Winn:2014xna}.

 For instance, \cite{Ford,AL,Juric} studied the origins of eccentric extrasolar planets through planet-planet scattering. Simulations with unequal-mass planets starting on nearly circular orbits resulted in broadening in the final distribution of eccentricities. Close encounters also increase the planet-planet scattering rate, enhancing dynamical instabilities and planetary ejection, see e.g.~\cite{2022MNRAS.515.5942B}. 
 
We also note that single transiting systems seemingly exhibit larger mean eccentricities, while multiples tend to have nearly circular orbits.   \cite{Limbach} identified a strong anticorrelation between the number of planets in a system and their orbital eccentricities. This may be indicative that highly eccentric systems are less stable, leading to ejection events \citep{2011MNRAS.411..859M}.


\section{The impact of Galactic Populations of PBH}
\label{sec4.0}

We next examine the plausibility that a galactic population of PBH could play a role in shaping the orbits of exoplanets. Specifically, we wish to obtain an estimate of the number of close encounters throughout the galaxy, as well as the typical relative velocity of the flyby. Subsequently, in Section \ref{sec7} we will proceed to examine the resulting deviations to eccentricities and semi-major axes of exoplanets due to PBH encounters in a simplified setting. 

Since neither observations nor simulations currently give complete or accurate predictions for initial values, we will consider a simplified system in which the galaxy is entirely comprised of Jupiter-star systems (later we also consider Neptune-star systems). Thus, each star has a single planet whose initial eccentricity and semi-major axis are roughly Jupiter-like with $(e_0,a_0)\sim$  (0, 5 AU).  Our aim is to identify the late time values of the orbital parameters 
\beq
e_\infty &=e_0+\Delta e\\
a_\infty &=a_0+\Delta a.
\eeq
 These perturbations $\Delta a$ and $\Delta e$ receive contributions from standard astrophysical bodies, such as close encounters with stars, as well as potentially from encounters with hypothetical bodies such as PBH. Moreover, one expects that $\Delta a$ and $\Delta e$ will depend on the radial distance from the galactic centre, since the densities and velocities of stars and PBH will change depending on the star's location in the galaxy.

\begin{figure}
\includegraphics[width=0.5\textwidth]{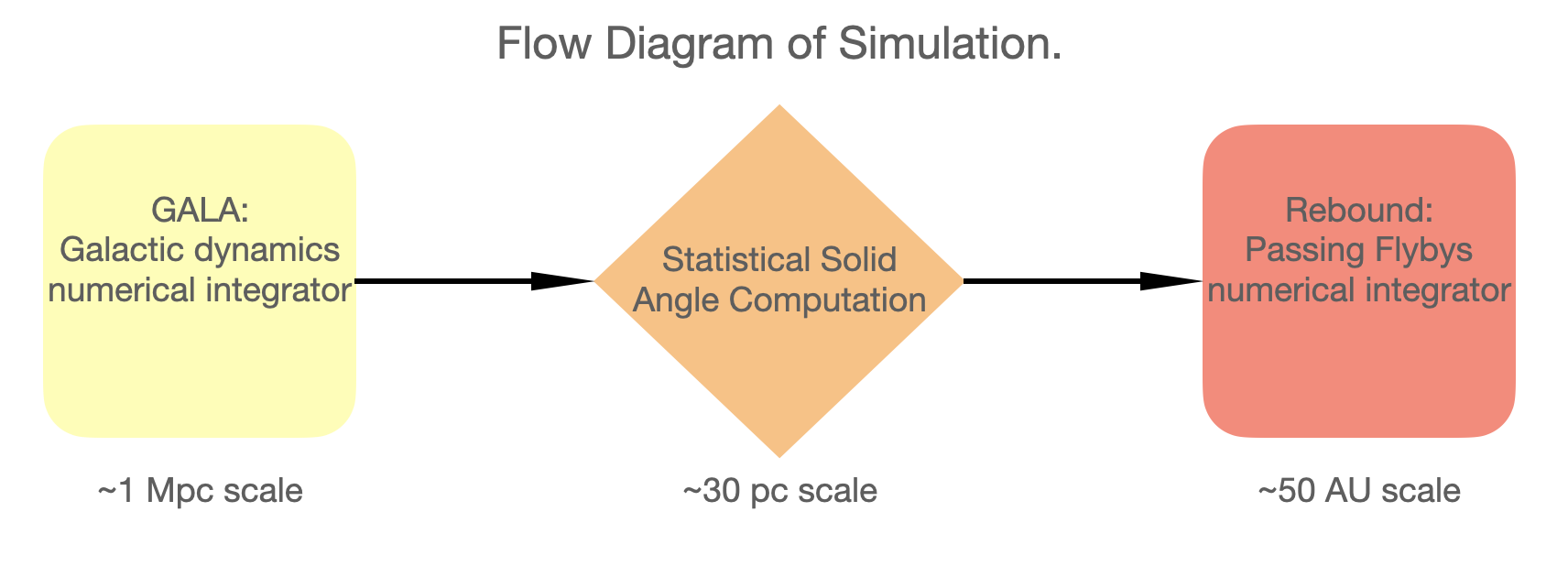}
\caption{~Flow diagram of the computations used to study how primordial black holes (PBHs) may impact exoplanet orbits.
\label{fig:flow-diagram}}
 \end{figure}

\subsection{Dynamics across 10 Orders of Magnitude}
\label{sec4}
  
  In modelling stellar close encounters, there is a good understanding of the typical number of encounters and the distribution of relative velocities of intruders \citep{Pfalzner,Cuello,Munoz,Winter}. However, for PBH, we will first need to calculate the typical likelihood of PBH close encounters for a given PBH population. Subsequently, in Section \ref{sec7} we will use this information to evaluate the impact of a population of PBH on exoplanet orbital parameters. We focus on the case that the intruding PBH passes through the system without being captured, i.e.~a one time `flyby'. While it is possible that the PBH could be captured (potentially leading to prolonged interactions between planets and the intruder), such events require significant energy dissipation via three-body interactions and are statistically substantially less likely to occur compared to flyby events.\footnote{We highlight the recent work of \cite{Lehmann:2022vdt} in relation to PBH capture by stars. However, this work does not consider the subsequent impact on planetary orbits.}

\begin{figure*}
    \centerline{
    \includegraphics[width=0.37 \textwidth]{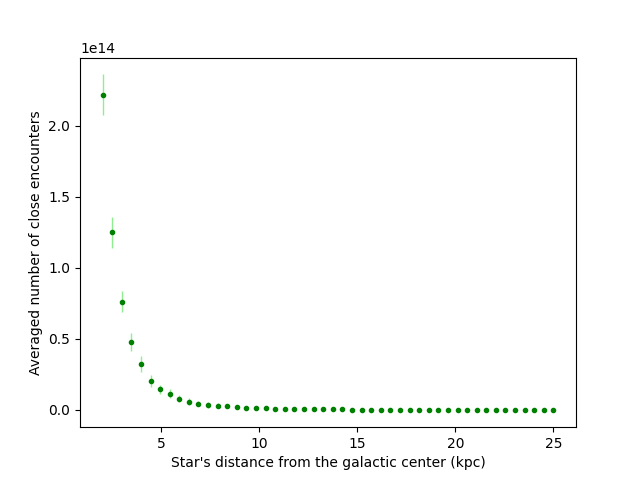} 
    \hspace{-7mm}
    \includegraphics[width=0.37\textwidth]{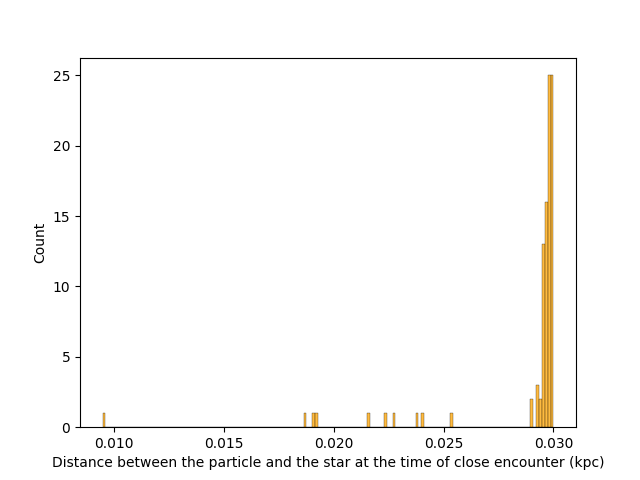}
    \hspace{-5mm}
    \includegraphics[width=0.37\textwidth]{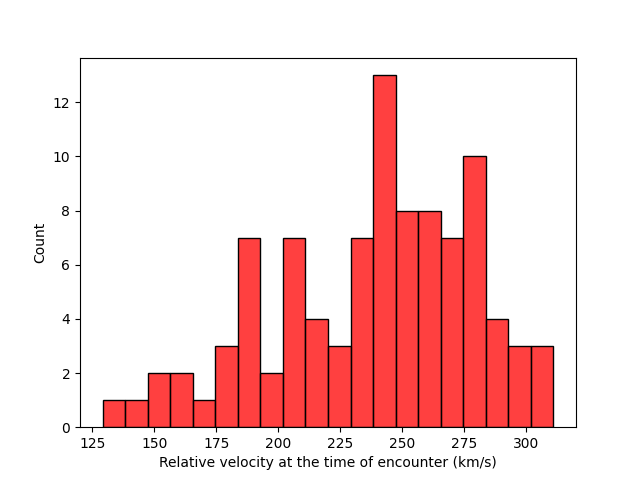}}
    \caption{Results of {\tt Gala} simulations. The star is assumed to have a circular orbit around the galactic center at a distance $d$; we consider $d=$2,3,4, and 5 kpc. The left panel shows the average number of close encounters as a function of the distance of the planet-star system from the galactic centre. The centre and right panels show histograms of the distance between the star and the intruder, and their relative velocities. The histograms show combined results for an equal number of runs at each value of $d$.}
    \label{fig:Distance_Distribution}
\end{figure*}

To study the potential impact of PBH flybys on exoplanet orbits, we break the problem into three parts: 
\begin{enumerate}
    \item[\bf A.] Given a star in a circular orbit around the galactic center at a given distance, how many PBH enter the star's local neighborhood within a given time period?
\vspace{3mm}
    \item[\bf B.] For a PBH that enters the neighborhood of a star, what is the probability that the PBH comes sufficiently close to appreciably perturb planetary orbits around the star? 
\vspace{3mm}
    \item[\bf C.] For a PBH that enters the planet perturbing region,  what is the statistical impact on the orbital parameters? 
\end{enumerate}  

By answering each of these questions in turn, we will explore whether PBH (or similar objects) can significantly impact the orbits of exoplanets. 
To make progress on these questions we will consider a simplified set-up of a galaxy outlined above, comprised solely of Sun-Jupiter systems.
Our aim is to provide a first look at how the statistical distribution of exoplanet orbits might be sculpted via PBH flybys. 

For part {\bf A}, to estimate the number of PBH that enter the neighborhood of a given star---which we take to be a sphere of radius $r_c=30$ pc---we implement numerical galactic dynamics simulations using the public \texttt{Gala} code \citep{Gala}. This Astropy-affiliated Python package numerically integrates the trajectories of stars and other astrophysical objects based on a given mass model. The computations are detailed in Section~\ref{sec:Gala}.

Then, for part {\bf B}, we analyze how many of these PBH that come within 30 pc (as defined by our simulation) lie on trajectories that would lead to a close encounter with a planet-star system. The distribution of distances is informed by the closest approaches found in our \texttt{Gala} simulations.  Thus, we estimate the likelihood that a given star would undergo a PBH flyby that could non-negligibly perturb its planetary system. We take the radius at which non-negligible flybys may occur to be 15 AU  of the star for Jupiter-like planets, and 90 AU for Neptune-like planets. Any PBH that comes within this distance we record and model in the next step. This calculation is detailed in Section \ref{sec:solid}.

Finally, for part {\bf C}, we use \texttt{REBOUND} simulations to sample how the exact details of the close encounters play out as detailed in Section \ref{sec7}. We use relative velocities informed by our \texttt{Gala} simulations and sample impact parameters for the PBH with $b < R$, then examine the flyby's effect on the planet's orbital parameters both in specific individual cases and as statistical distributions. 

This three-step subdivision is necessary to make the problem tractable. The scale of the galaxy is $\sim1$ Mpc, while stellar systems are of order $10^{-4}$ parsec (this is the Neptune-Sun distance), thus the problem spans 10 orders of magnitude in distances. Compartmentalizing the problem into three units tracks the relevant dynamics first at the galactic scale using \texttt{Gala}, then at the parsec scale via solid angle scaling calculations, and, finally, at the interplanetary scale $\lesssim50$ AU using \texttt{REBOUND}. Figure~\ref{fig:flow-diagram} shows the flow diagram of our approach.

\vspace{-2mm}
 \subsection{Simulating Galactic Dynamics}
\label{sec:Gala}

We implement galactic simulations using  \texttt{Gala} to determine the frequency of PBHs entering the neighborhood of a given star.  The code models the PBHs and a given star as test particles and traces out the trajectories of these test particle. We use the \texttt{MilkyWayPotential} model, consisting of a spherical nucleus and bulge, a Miyamoto-Nagai disk \citep{1975PASJ...27..533M}, and a spherical dark matter halo. Given that the stars and PBH are all treated as test particles, the results of our galactic simulations do not depend on their masses.

Since modeling a galactic population of PBH is computationally intensive, we simplify the analysis by analyzing a single star at a time (at some radius $r$ from the galactic center) and model only the PBH that are sufficiently nearby that they could potentially have a close encounter within 1 Myr. 
A PBH on a circular orbit at 2 kpc with velocity 220 km/s travels $0.225$ kpc in 1 Myr, thus it traverses $1.8\%$ of the full orbit at 2 kpc, or an angular displacement of $\theta_*\approx0.11$ rads.
We consider a radial region of the galaxy such that the star of interest is at $\theta=0$ and select a galactic sector (or ``slice'') defined by $\theta\in(\theta_-,\theta_+)$ with $\theta_\pm=\pm\theta_*$.
Within this galactic sector, we randomly distribute 50,000 test particles sampling from an NFW density profile \citep{Navarro:1996gj} to act as PBH.
Each PBH is assigned an initial speed of 220 km/s with a randomized three-dimensional direction. To represent the star, we introduce a test particle at a distance between 2 kpc and 25 kpc from the galactic center, with a circular velocity of 100 km/s.

 \texttt{Gala} calculates the dynamical evolution through a background potential in which all of the test particles move. We utilyse the \texttt{Leapfrogintegrator}, a symplectic integrator that computes the position coordinates and velocity vectors of particles with specified timesteps.  For each time step, we evaluate the pairwise distances between the target star and each of the PBH.
We identify the number of objects that pass within  $r_c = 30$ pc of the star and record the distance and relative velocity of each object on the first time-step that they are within the ball of radius $r_c$ around the specified star. We use these distributions of initial positions and velocities in subsequent stages of our analyses. 

While we integrate the {\tt Gala} simulations for 1~Myr, the age of a typical star (and the Milky Way) is roughly 10 Gyr. 
Thus, to get a final estimate of the number of close encounters, we scale the number of encounters by a factor of $10^4$ to adjust for this. 
The final rescaled count of the number of bodies passing within 30 pc of the star, averaged over 72 runs of our {\tt Gala} simulation, is given in Figure \ref{fig:Distance_Distribution} (left) alongside the position-velocity distributions of the PBH.

\begin{figure}
        \includegraphics[width=0.38\textwidth]{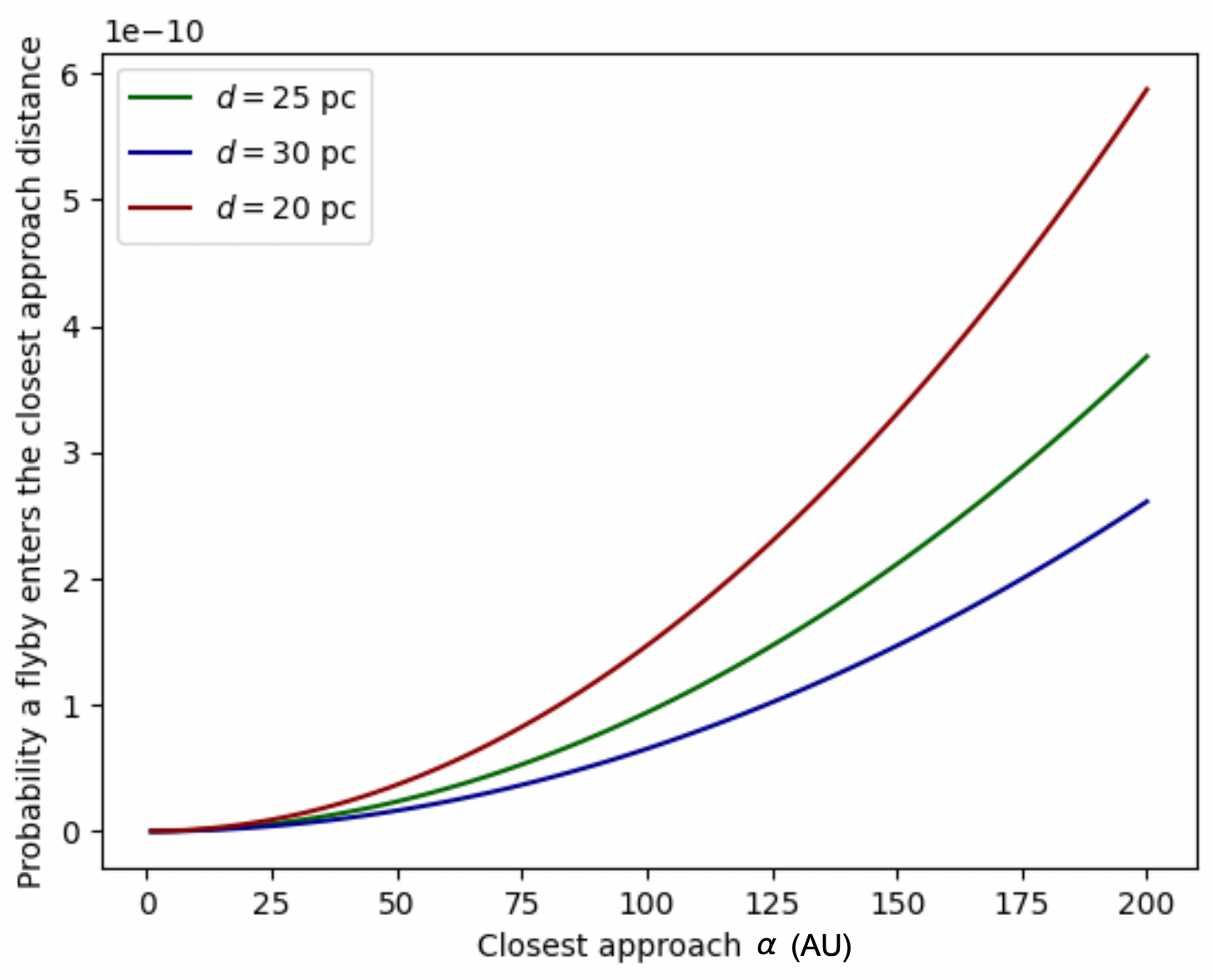}
\caption{The probability that a PBH enters a given closest approach distance $\alpha$ from an initial distance $d$. For star-Jupiter (star-Neptune) systems we take $\alpha=15$ AU (90 AU).
 \label{fig:solid-angle-L}
 \vspace{1mm}
}
\end{figure}

Note that we have assumed that this sector of the galaxy has 50,000  PBH, this implies the total PBH population in the galaxy is $N_{\rm PBH}\sim10^{6}$. The total mass of dark matter in Milky Way inferred by GAIA is $10^{12}M_\odot$ \citep{Cautun:2019eaf}.  Suppose that the PBH have a common mass of $10^{-1}M_\odot$, then these PBH $\Omega_{\rm PBH}$ constitute a tiny fraction of the observed dark matter relic abundance $\Omega_{\rm DM}$
\beq
f_{\rm PBH}\equiv\frac{\Omega_{\rm PBH}}{\Omega_{\rm DM}}\sim \frac{N_{\rm PBH} M_{\rm PBH}}{10^{12}M_\odot}\sim 10^{-7}\left(\frac{M_{\rm PBH}}{10^{-1}M_\odot}\right).
\label{eq5}\eeq
Thus, the inferred dark matter relic abundance $\Omega_{\rm DM}\approx0.26$ is assumed to be dominantly in the form of some new particle dark matter species, with an insignificant contribution due to PBH ($\Omega_{\rm PBH}\sim10^{-8}$). 

For substellar mass PBH, observational constraints on PBH (e.g.~microlensing) are satisfied for sub-percent level abundances $f_{\rm PBH}\lesssim10^{-2}$ \citep{Carr:2020xqk}. Additionally, if dark matter can annihilate to Standard Model states, then this can lead to limits, in particular, from observations of extragalactic $\gamma$-rays \citep{Fermi-LAT:2015qzw}. For $f_{\rm PBH}\sim10^{-7}$ such indirect detection bounds exclude thermal (``WIMP'') dark matter with velocity-independent annihilation cross sections \citep{Adamek:2019gns,Lacki:2010zf,Boucenna:2017ghj,Gines:2022qzy,Chanda:2022hls}, however, other dark matter particles remain viable. Examples of viable dark matter scenarios include: velocity-dependent (``p-wave'') annihilations, freeze-in dark matter \citep{Hall:2009bx,Elahi:2014fsa} (for limits see \cite{Gines:2022qzy,Kadota:2022cij,Chanda:2022hls,Chanda:2025bpl}), and (non-thermal) gravitationally coupled dark matter.

Since {\tt Gala} treats the PBH as test particles, the only impact on varying $M_{\rm PBH}$ is that it changes our assumption regarding $f_{\rm PBH}$. 
However, at later stages of our analysis involving simulations of the close encounter varying $M_{\rm PBH}$ will significantly alter the results. We note that it is a common expectation for PBH to have a tight mass spectrum due to the details of their primordial production mechanism \citep{Carr:2020xqk}. The assumption that the PBH have equal mass is typically considered a reasonable approximation.


\begin{figure}
            \includegraphics[width=0.4\textwidth]{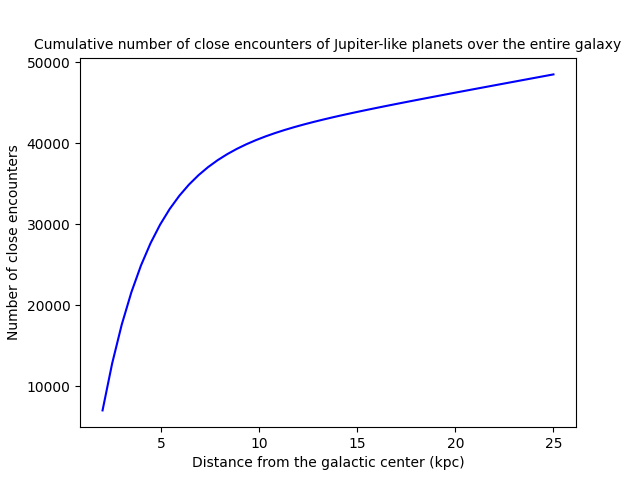}
    \caption{
For $3\times10^6$ PBHs over the entire galaxy, we show the number of systems that undergo PBH flybys (taken to be within a radius $\alpha=15$ AU). This is shown as a cumulative count in terms of increasing distance to the galactic centre. The initial velocity vectors and distance from the star are sampled from outcomes of our Gala simulations.  The results are averaged over $10^5$ simulations on a sector of the galaxy and then scaled to the full galaxy.
\vspace{1mm}
 \label{fig:solid-angle-R}}
\end{figure}


\subsection{Probability of Close Encounter}\label{sec:solid}

The above {\tt Gala}  analysis estimates the frequency of PBHs coming within 30 pc of the star. While close in galactic terms, this is huge distance relative to solar system distances (nb.~1pc$~\sim10^5$AU). In the second step of our analysis, we analyze the probability that a PBH within this  30 pc radius actually undergoes a close encounter with the planet-star system, which we take to mean comes within $\mathcal{O}(10)$ AU.

We compute whether a given PBH in the neighborhood of a star ($d < r_c = 30$ pc) has the appropriate solid angle region $\Omega_d$ such that its perturbation of planetary orbits is non-negligible. We require the PBH to pass within 15 AU of the star for Jupiter-like planets or 90 AU for Neptune-like planets.  The difference in threshold corresponds to the change in semi-major axes (nb.~$a_J\sim5$~AU and $a_N\sim30$~AU).

We calculate the solid angle $\Omega_d$ defined by the area that a ball of radius $\alpha$  ($\sim \mathcal{O}(10)$ AU) centered on the star intersects the ball of radius $d$  ($\sim \mathcal{O}(10)$ pc) centered on the intruder.  The probability $p_d$ that a PBH which starts at an initial distance $d$ enters a ball around the star of radius $r$ is related to $\Omega_d$ as follows
\beq
p_d=\frac{1}{4\pi}\Omega_d\approx\frac{1}{4}\left(\frac{\alpha}{d}\right)^2. 
\eeq
For example, the probability that a flyby at initial distance 30 pc from the star passes within 15 AU of the star is roughly $p\sim p_{r_c}\sim10^{-12}$. 
Figure \ref{fig:solid-angle-L}  shows the probability that a PBH at initial distance $d=20,25,30$ pc passes within a given distance of closest approach ($\alpha$) of a star. 

Thus we can calculate the number of PBHs entering the region 15 AU around the star, by rescaling our previous {\tt Gala} plot (Figure \ref{fig:Distance_Distribution} (left)) by a factor of $p_d$ where $d$ is sampled over the {\tt Gala} outputs (Figure \ref{fig:Distance_Distribution} (centre)).  
For each computation, the PBH's initial distance from the star ($d < r_c = 30$ pc) is randomly drawn from a distribution of distances at the time of close encounter from our {\tt Gala} simulations (cf.~Figure \ref{fig:Distance_Distribution}). 
Finally, since we are currently considering a sector of the galaxy, we need to rescale our results to the whole galaxy.  Then, we arrive at an estimate for the expected number of star-planet systems that experience a sufficiently close encounter with a PBH to potentially alter their orbital parameters. Figure \ref{fig:solid-angle-R} shows the number of stars that undergo a close encounter with a PBH in the galaxy (with our assumptions), showing a cumulative count as the stars' distance from the galactic center is increased.

\vspace{1mm}
\section{Distributions of Exoplanet Orbits Following Encounters}
\vspace{1mm}
\label{sec7}

In the previous section, we have endeavored to estimate how many stars may undergo close encounters with a PBH. In this section, we explore the implications for those planet-star systems that do  encounter a PBH flyby. 
We will first simulate a range of impacts due to PBH flybys and then examine the statistical changes to planetary orbital parameters due to close encounters of exoplanet systems with PBH.

We simulate the flybys using \texttt{REBOUND}, as explained in Section \ref{sec:REBOUND}.
We use a hybrid integration scheme which switches from \texttt{WHCKL} \citep{ReinTamayoBrown2019} to \texttt{IAS15} \citep{IAS15} and back to \texttt{WHCKL} if the flyby object passes within $30a$, thirty times the semi-major axis of the planet (as discussed in Section \ref{sec:REBOUND}). 
The PBH flyby paths are introduced via hyperbolic trajectories defined by their impact parameter  $b$. 
As previously, to simplify the analysis, we assume all exoplanets are in simple planet-star systems with initially circular orbits.
 In our primary case of interest the planets have mass $10^{-3}M_\odot$ and initial orbital parameters $(e_0,a_0)\sim$  (0, 5 AU) -- ``Jupiter-like''. 
 We also consider Neptune-like planets with mass $5\times10^{-5}M_\odot$ and $(e_0,a_0)\sim$  (0, 30 AU).  This starting configuration is clearly a simplification, in particular, since we expect formation histories to impact $a_0$ and $e_0$, and also one anticipates close encounters with stars for a significant fraction of systems (we discuss this further in Section \ref{limit}). In Appendix \ref{Ap1} we briefly examine how varying the parameters  $a_0$ and $e_0$ alter the results.

\begin{figure*}
    \centering
    \includegraphics[width=0.67\textwidth]{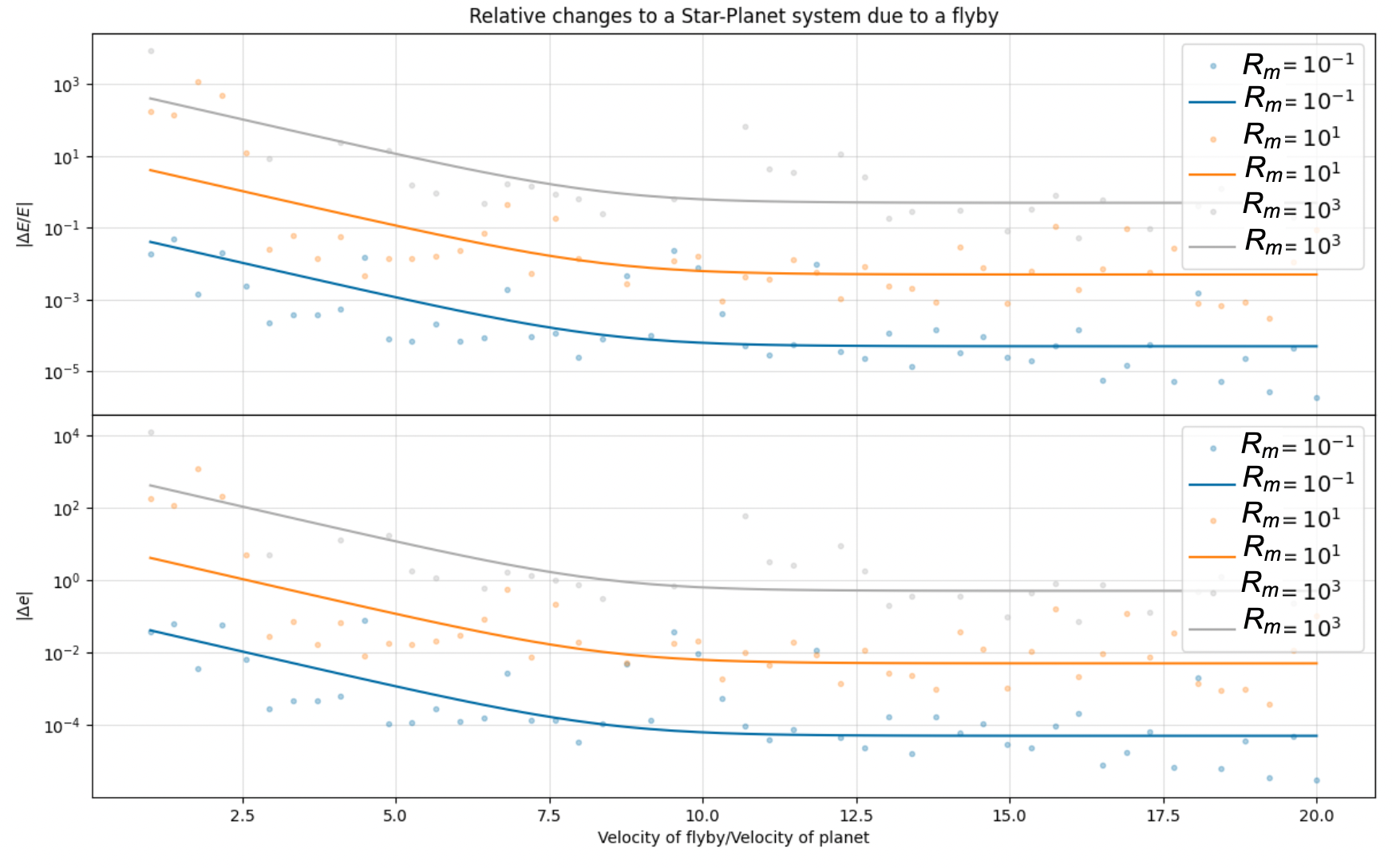}
\vspace{-3mm} 
 \caption{{\bf Changing the velocity of the intruder.} Changes in the relative energy and eccentricity of the planet due to encounters with passing flybys at impact parameter $b=10^{-4}$ as the ratio of the flyby's velocity to that of the planet is varied. We show our results for different fixed values of $M_{\rm PBH}$ written in terms of $R_{M}\equiv M_{\rm PBH}/M_p$ as shown, where $M_p$ is the mass of the planet. }
    \label{fig:flyby_main}
    \includegraphics[width=0.67\textwidth]{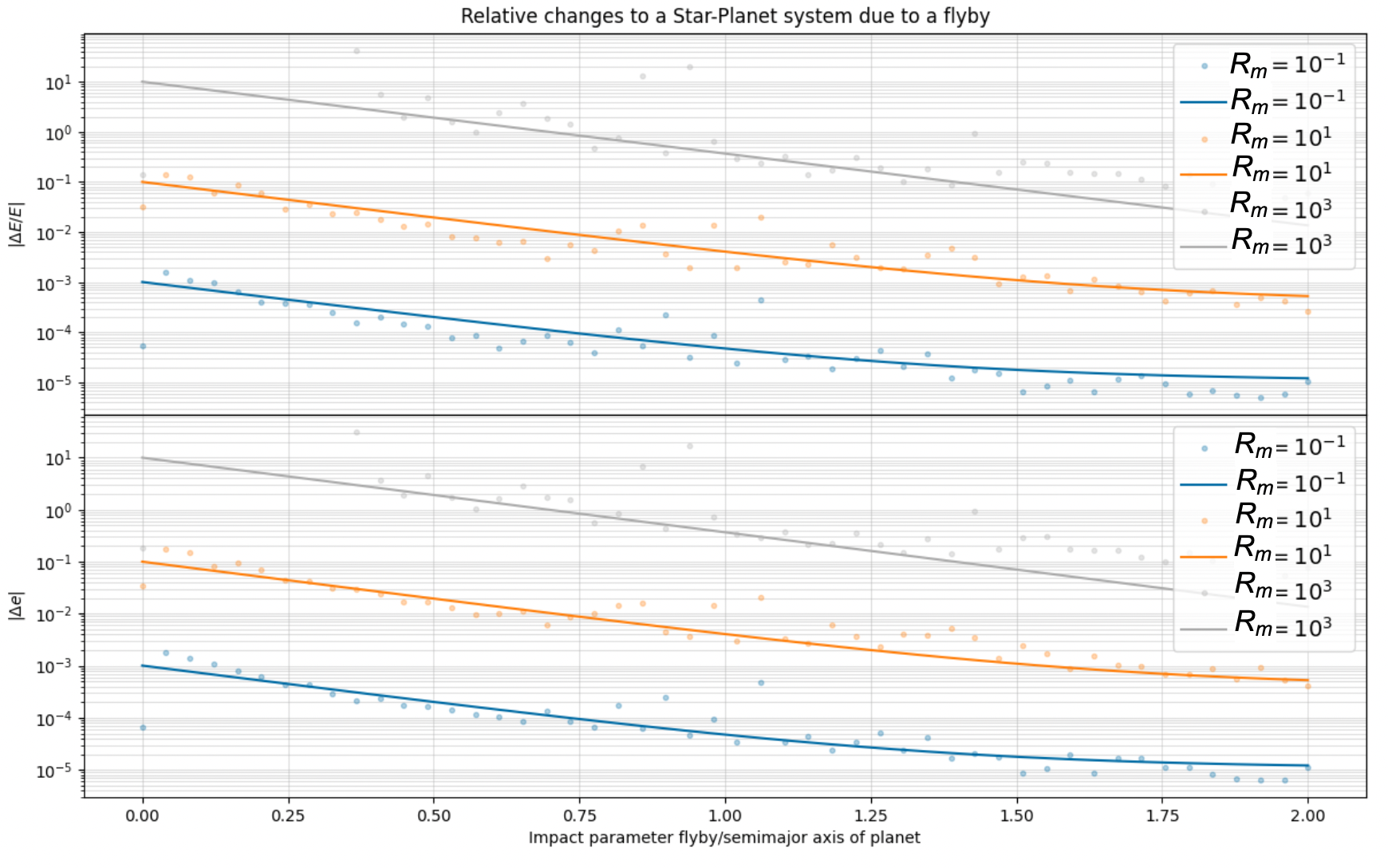} 
\vspace{-3mm} 
    \caption{{\bf Changing the impact factor.} Changes in the relative energy and eccentricity of the planet due to encounters with passing flybys where we vary the  impact parameter $b$  relative to the planets semi-major axis $a$. The flyby's velocity is fixed to be 200 km/s. }
    \label{fig:flyby_impact}
    \centerline{  \includegraphics[width=0.4\textwidth]{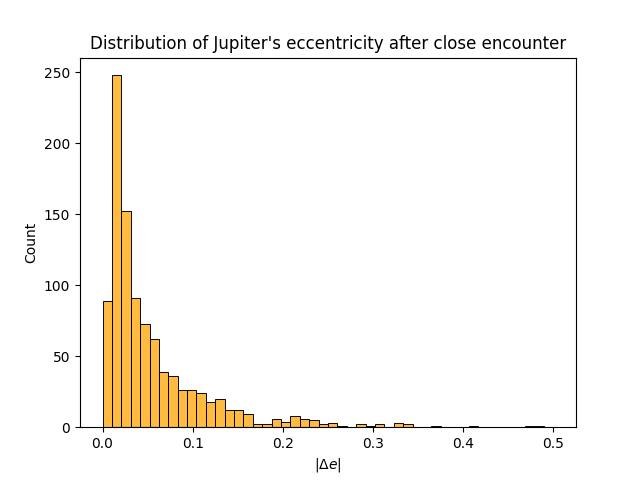}
      \includegraphics[width=0.4\textwidth]{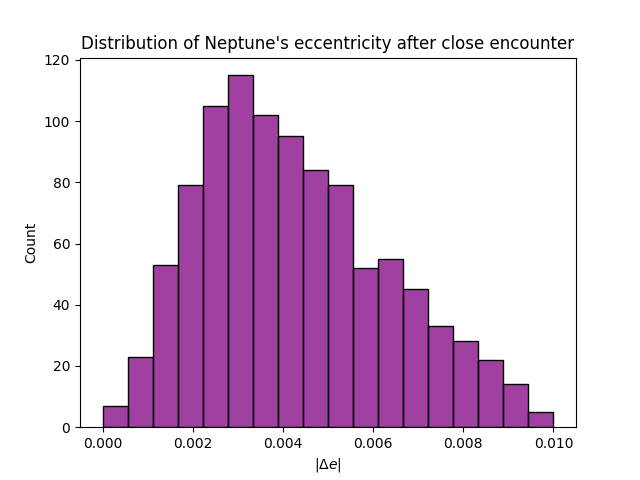}}
    \caption{The distribution of the resulting eccentricities of Jupiter (left)/Neptune (right) planet-star systems after a close encounter with a flyby PBH of mass $10^{-1} M_\odot$ and velocity $200$ km/s (taking for Jupiter $b_J\in 15\sqrt{U[0, 1]}$AU and for Neptune $b_N\in 90\sqrt{U[0, 1]}$AU). 
      \label{Fig:jupiter-neptune-dist}}
\end{figure*}

We shall first examine the impact on the late-time orbits of the planet from varying the parameters of the PBH. Figure \ref{fig:flyby_main} shows the relative change in energy ($ \Delta E/E\propto \Delta a$) and change in eccentricity from zero of the planet due a passing flyby at small impact parameter, taking $b=10^{-4}$. We consider six fixed values of $R_M=M_{\rm PBH}/M_p$, where $M_p$ is the mass of the planet.
On the $x$-axis we vary the intruder velocity, in terms of $R_{v}$ the ratio of the flyby speed relative to the orbital speed of the planet $R_v\equiv |v_{\rm PBH}|/|v_p|$. Each point in Figure \ref{fig:flyby_main} corresponds to the average over 10 simulations. Note that for different values of $R_M$, the general shapes of the curves are highly similar.

The perturbation strength decreases exponentially with respect to increases in the flyby velocity until the threshold value $R_{v} \approx 10$, whereafter the perturbation strength is insensitive to further increases in velocity. For a Sun-Jupiter system, an intruder of mass $M_I = 10^{-2} M_\odot$ and velocity $|v_0| \approx 130$ km/s can lead to percent-level changes in planetary eccentricities ($\Delta e \sim 0.01$). 

In the subsequent studies detailed below, to simplify the analyses, we fix the relative speed of all encounters to be 200 km/s. 
This choice is informed by our {\tt Gala} results (Figure \ref{fig:Distance_Distribution}), which found the range to lie between 125 km/s and 300 km/s. 
Notably, from consideration of Figure \ref{fig:flyby_main}, one can see that the results are relatively insensitive for higher relative velocities (i.e.~over this range of interest), thus this simplifying assumption is reasonable. 
Furthermore, we obtained these results with relative velocities also fixed to 100 km/s, and these were not qualitatively different.

Figure  \ref{fig:flyby_main} assumes the PBH comes very close to the star. 
Thus, in Figure \ref{fig:flyby_impact} we examine how these results vary as we move away from near-zero impact factor. Increasing the impact factor corresponds to a more distant point of closest approach for the flyby, cf.~eq.~(\ref{rp}). As can be observed in Figure \ref{fig:flyby_impact}, and may be expected, closer flybys lead to more significant orbital changes.  

The final eccentricity is determined by $e_f = e_0 + \Delta e$, with our stated assumption $e_0$ this implies $e_f = \Delta e$. Notably, for close encounters with $e_f> 1$, the planet will become unbound from the parent star. An interesting consequence of this is that such close encounters could lead to an excess of free-floating planets above those expected due to stellar close encounters.  Interestingly, the OGLE telescope \citep{Mroz:2017mvf} reported a tentative excess of microlensing events corresponding to Earth-mass bodies. There is also an apparent excess of free-floating planets in the Upper Scorpius young stellar association \citep{2022NatAs...6...89M}. The OGLE excess was interpreted as an abundance of PBH \citep{Niikura:2019kqi,Scholtz:2019csj}, but it could also be induced by an excess in free-floating planets (potentially arising from close encounters between PBH and stellar systems). Upcoming observations by the Rubin Observatory and Roman Space Telescope should provide new insights into populations of sub-stellar mass free-floating objects, with significant discovery potential.

Finally, we next turn to part {\bf C} of Section \ref{sec4.0} and examine how PBH flybys may alter the statistical distribution of exoplanet orbits.
For each set of initial orbital parameters, we implement a Monte Carlo sampling technique to randomly choose the parameters of the flyby. 
For star-planet systems, the impact parameter $b$ of the flyby dictates the distance of closest approach and is drawn from a random distribution such that  $b_i=r_i \times \sqrt{ U[0,1] }$, where $r_i$ is the radius of critical interaction distance, and $U[0,1]$ is a random variable from a uniform distribution between 0 and 1. For star-Jupiter systems $r_J=15~{\rm AU}$, whereas for star-Neptune systems $r_{N}=90$ AU. 
The angular parameters $\Omega$, $\varpi$, and inclination $i$ of the flyby are all uniformly drawn from $[-\pi, \pi]$. Moreover, each initial setup is ran for $10$ samples, with each run independently sampling the angular orbital parameters. 

Our findings are presented in Figure \ref{Fig:jupiter-neptune-dist}, which shows the distribution of changes to the eccentricity for Jupiter and Neptune-like planets assuming a PBH intruder with mass $10^{-1}M_\odot$. Changes to Jupiter's eccentricity are strongly peaked at small values, whereas Neptune has a broader distribution. We attribute these differences to the fact that Jupiter is twenty times more massive than Neptune.

\clearpage


\section{Limitations}
\label{limit}

Modelling PBH encounters in the galaxy is rather complicated, not least because the abundance of PBH is unknown.\footnote{The spatial distribution is also uncertain, as dynamical friction is expected to lead to deviations away from NFW for the PBHs.}  Moreover, the mass spectrum of PBH is largely unconstrained. Even assuming a uniform mass $M_{\rm PBH}$, a wide range of $M_{\rm PBH}$ are permissible. Varying $f_{\rm PBH}$ or $M_{\rm PBH}$ can significantly alter the results. 
We have taken fixed values for  $N_{\rm PBH}$ and $M_{\rm PBH}$ to be able to proceed. We have also had to make several simplifying assumptions, as we discuss below. We reiterate that our aim is to give an appreciation of the fact that exoplanet orbits could be altered by a galactic population of  PBH and not to extract precise predictions.

One major simplification in this work is that we have modeled our galaxy by assuming that each star is a two-body system with a 1$M_\odot$ star and a single planet which we take to be either Jupiter or Neptune-like. We have ignored the existence of binary star systems, as well as the rich variety of planetary systems. Appendix \ref{Ap1} makes a brief study on varying the mass and semi-major axis of the planet, but the true range of planetary systems is, of course, much greater. 
Let us start with the assumption that the parent star is Sun-like (i.e.~1$M_\odot$). On the low mass side, planets have been found orbiting brown dwarfs\footnote{Planets also orbit neutron stars (e.g.~PSR B0943+10 with 0.02$M_\odot$ \citep{Suleymanova}), but the orbits of any surviving planets would be strongly impacted by the supernova.}  with masses 0.08$M_\odot$ to 0.13$M_\odot$. At the other extreme, a planet has been observed in orbit around the 6$M_\odot$ star b Centauri (HD 129116). 
While single planet systems are thought to be common (this conclusion may be biased by observational capabilities), multi-planet systems are far from atypical.  It has been estimated that multi-planet systems comprise around a quarter of all Kepler planets \citep{Sandford}. Kepler-90 (a Sun-like star 2.8 light years away) has eight planets \citep{Shallue}.  Even for single planet systems the planetary mass can range from $\sim10^{-2}M_\odot$ (i.e.~10 $M_J$)\footnote{At masses over $0.013M_\odot$ thermonuclear fusion to deuterium is typically possible. Although \cite{exoplanets} includes entries up to $0.08M_\odot$, the bodies above $0.013M_\odot$ are likely Brown Dwarfs.}, for instance ROXs~42Bb \citep{Currie}, down to $\sim10^{-7}M_\odot$ being the mass of both Mercury and PSR~B1257+12A \citep{Konacki}.

We also do not consider the possibility of PBH capture or multiple PBH flybys (by different PBH). 
For PBH capture, on the one hand, we expect capture events to be rare. However, a captured body would lead to a prolonged period of gravitational interactions between the intruder and the exoplanets, implying more significant disruption of the system. We highlight two recent papers on PBH capture \citep{Lehmann:2020yxb,Lehmann:2022vdt}. Moreover, it has been suggested that the evidence for the Planet 9 hypothesis \citep{Batygin} could be interpreted as an Earth mass PBH which was gravitationally captured in our Solar System \citep{Scholtz:2019csj}. We comment on the likelihood of multiple PBH encounters in Appendix \ref{multi} using simple analytic estimates to argue that this is negligible for small PBH fractional abundances, while for larger PBH fractional abundances ($\gtrsim$1\%) multiple flybys may be relatively common.

Furthermore, and related to multiple flybys, it is anticipated that many stellar systems will undergo close encounters with other stars and binary star pairs, both in star-forming regions and in the galactic field \citep{2015MNRAS.448..344L}. Thus, even starting from our assumed initial uniform distribution $(e_0,a_0)\sim$  (0, 5 AU), these stellar close encounters will imply a non-trivial smearing in the distributions of eccentricities and semi-major axes prior to (or in the absence of) PBH encounters.
Since stellar encounters are known to occur, we should endeavour to include this in our modelling. This could be incorporated by considering the distribution of the orbital parameters assuming a possible initial encounter with another star, leading to some non-uniform distribution of $e$ and $a$. Previous analyses, e.g.~\cite{2009ApJ...697..458S}, have examined the distribution of orbital parameters after a stellar flyby, however, the results are not in a form we can immediately adapt for our current implementations.

One could proceed by assuming that most stellar close encounters occur in stellar birthing regions and use {\tt Rebound} to calculate the distributions of $\Delta e$ following such encounters. This would lead to a set of histograms for eccentricity and semi-major axes analogous to Figure \ref{Fig:jupiter-neptune-dist}. Stars can have a rather different impact due to their velocities being generally much lower than a typical PBH. We would then rerun the analysis of Section \ref{sec7}, but here, rather than a uniform initial distribution $(e_0,a_0)\sim$  (0, 5 AU), we use the resulting histograms found from our {\tt Rebound} simulations of stellar encounters in the birthing cluster. This study, while interesting, is beyond the scope of this initial examination, but we plan to return to these questions in future work.

We conclude this section by noting that while our models clearly present a simplified picture of the rich dynamics of the galaxy, we believe that this work provides a reasonable starting point to discuss the potential for PBH to influence the orbits of exoplanets. In particular, these assumptions were critical for us to make progress towards such questions using reasonable computational resources. 

\section{Concluding Remarks}
\label{conc}

The orbits of planetary systems are established through a complex interplay of proto-planetary disk evolution, planet-planet interactions, and close encounters. If a PBH passes sufficiently close, it can potentially perturb planetary orbits similar to an intruding star.  Here we have made the first investigation into whether PBH flybys may also be involved in this complex history, and argued that they could play an appreciable role. The magnitude of their impact depends critically on the mass scale and abundance of the galactic population of PBH. 

In this work we have utilysed the public package {\tt Gala} to quantify the expected the number of PBH flybys for a given abundance and have shown that even with a small abundance ($f_{\rm PBH}\sim10^{-7}$) a significant number of stars would have undergone a close encounter with a PBH. We then implemented a Monte Carlo study using a modified version of the public code {\tt Rebound} to quantify how significantly the orbits would be perturbed. We leave for future work the interesting and feasible prospect of multiple PBH encounters on a system, and the prospect of a stellar encounter followed by PBH encounters. We would also like to examine more complicated stellar systems beyond the simplified Jupiter-Sun systems studied here.

It is interesting to consider whether a population of exotic bodies may be able to explain variations or anomalies in the orbits of planetary systems. In principle, precision measurements of exoplanet orbital parameters could be used to infer or constrain the abundances of PBH; however, in practice, the large uncertainties relating to both measurements and planetary formation present significant obstacles. In the optimistic case that future precision observations and modelling of exoplanet orbital parameters were completed, this could potentially provide a new class of `dynamical' constraints on PBH (cf.~\cite{Carr:1999,Carr:2020gox}). 

While we do not expect to be able to use exoplanet observations to place constraints in the near future, this work outlines the general principles of how one might use a future precision catalogue of exoplanets to discover or constrain populations of PBH. Finally, we note that if one could identify a class of clean planet-star systems that always formed with well-defined orbital parameters, these could provide an ideal testing ground for such dynamical constraints on PBH. Intriguingly, hot Jupiters seem to typically be found in single-planet systems \citep{Wright,Steffen,Steffen2}, although exceptions have been observed; for instance, WASP-47b \citep{Becker}. Another potential route towards discriminatory power could be to look for objects with inclined orbits; however, even in our own solar system, we do not fully understand the origins of planetary orbits away from their ecliptic plane (see e.g.~\cite{Batygin}). We leave these directions for future work.

 \section*{Acknowledgements}
JU is supported by NSF grant PHY-2209998 and thanks the Berkeley Center for Theoretical Physics, Queen's College, Oxford, and Rudolf Peierls Centre for Theoretical Physics, of the University of Oxford for their hospitality during this work. LH thanks the MIT-PRIMES program. We are grateful to Hanno Rein and Jakub~Scholtz for helpful discussions and the anonymous referee for their useful remarks.

\appendix


\setcounter{figure}{9}
\setcounter{table}{1}

\section{Varying Parameters}
\label{Ap1}

\begin{figure*}
    \centering
 \centerline{       \includegraphics[width=0.75 \textwidth]{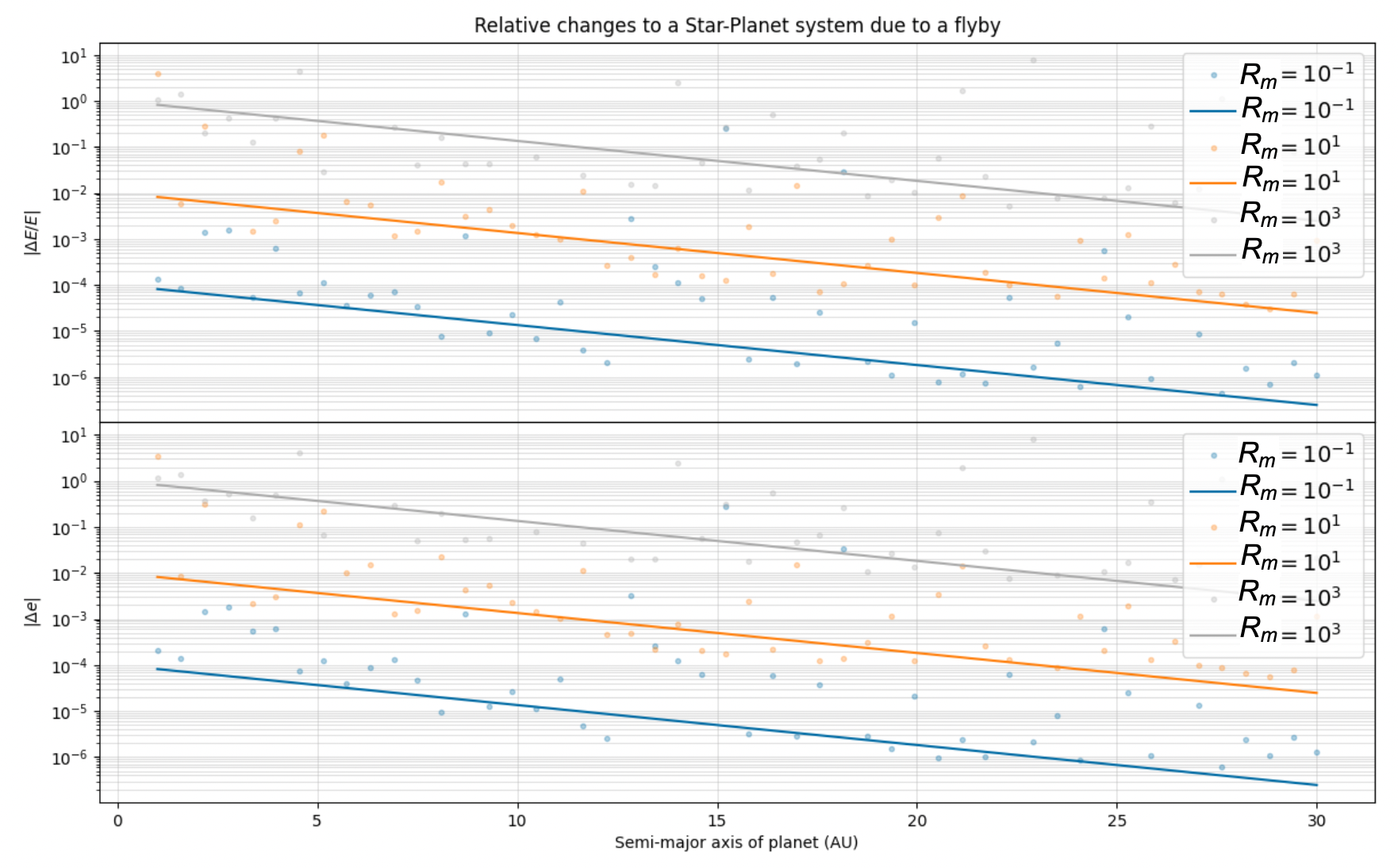}}
    \caption{{\bf Changing the initial semi-major axis of the planet.} Changes in the relative energy and eccentricity of the planet due to encounters with passing flybys with initial speed $|v_0| \approx 200$ km/s with the ratio of flyby's mass to the planet mass being $R_M = 10^{-3}$, $10^{-1}$, and $10^{1}$. Each point is the average of 10 simulations.     \label{fig:varying-planet2}}
\end{figure*}

\begin{figure*}
  \centerline{    \includegraphics[width=0.75\textwidth]{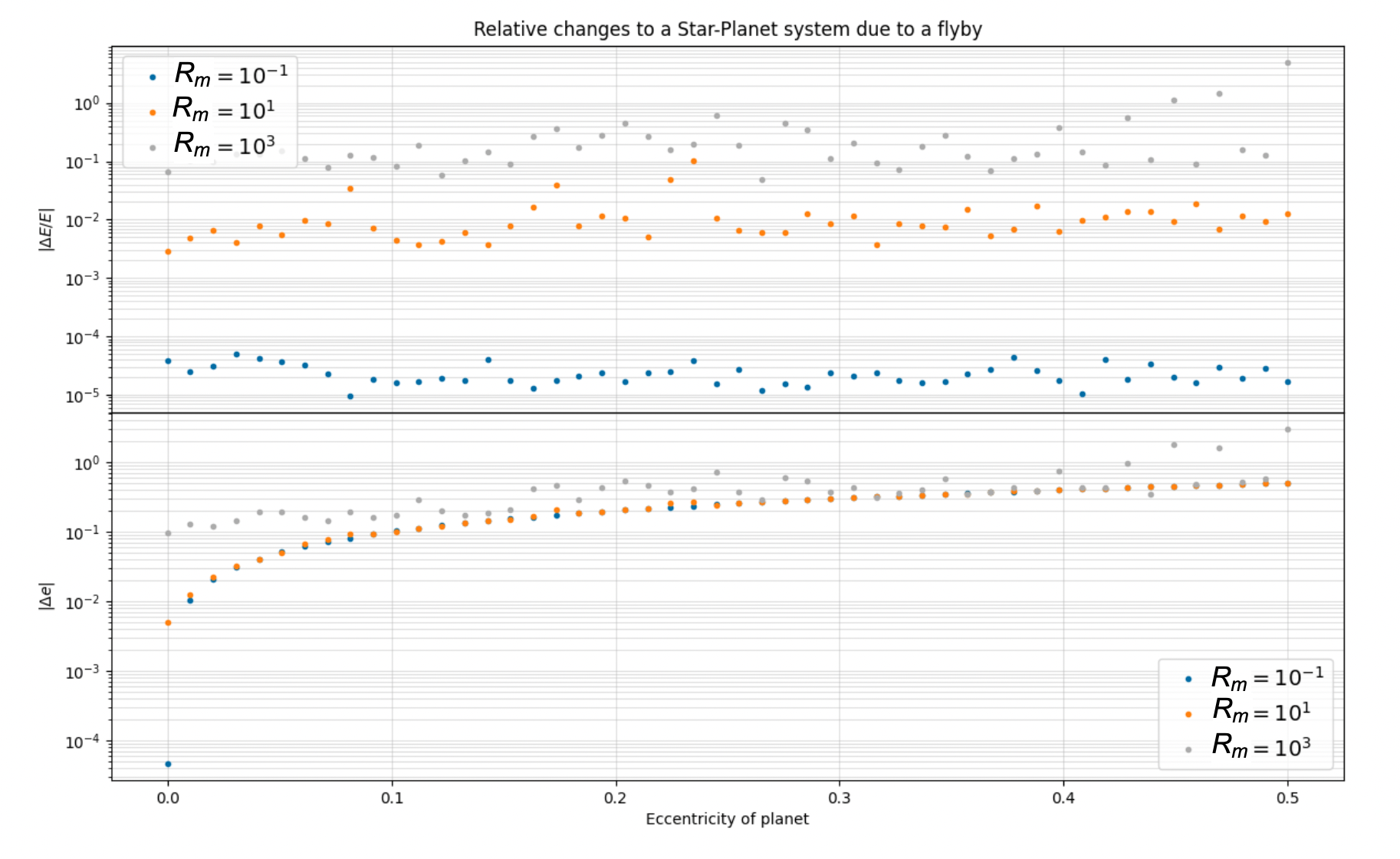}}
    \caption{{\bf Changing the initial eccentricity of the planet.} Changes in the relative energy and eccentricity of the planet due to encounters with passing flybys with initial speed $|v_0| \approx 200$ km/s with the ratio of flyby's mass to the planet mass being $R_M = 10^{-3}$, $10^{-1}$, and $10^{1}$. Each point is the average of 10 simulations.   \label{fig:varying-planet1}}
\end{figure*}

In Section \ref{limit} we discussed the rich variety of stellar systems and contrasted this to our assumption that the stars of the Milky Way could be modelled as Sun-Jupiter systems. In this appendix, we would like to provide a flavour for how changes to orbital parameters vary with respect to changes in the assumption of the model.   Taking a fixed flyby speed $|v_0|=200$ km/s and impact factor $b=10^{-4}$, Figures \ref{fig:varying-planet2} and \ref{fig:varying-planet1} indicate the changes to the orbital parameters as a function of the initial eccentricity and semi-major axis.

\section{Multiple PBH Encounters} 
\label{multi}

In this appendix, we comment on the prospect that systems may have multiple encounters with PBH, in which case the orbital parameters would evolve with successive flybys. 
While a full study is beyond the scope of this work, an immediate question one might ask is how likely are multiple encounters with PBH? We can use a similar Poisson distribution estimate as applied to the stellar case in Section \ref{sec3}. The rate of PBH encounters can be expressed as follows
\beq
\Gamma_\bullet t_g
&\simeq n_{\rm PBH}\sigma v_{\rm PBH}  t_g \\
&\sim
\left(f_{\rm PBH} \frac{\rho_{\rm DM}}{M_{\rm PBH}} \right) \sigma v_{\rm PBH}  t_g~,
\eeq 
where we have used that the PBH number density is related to the PBH fractional abundance in terms of the local dark matter abundance $\rho_{\rm DM}$. 
The target star system is the same as the stellar case, so we take $\sigma=\pi (1000~{\rm AU})^2$, the velocity of PBH we take to be $v_{\rm PBH}\simeq v_{\rm DM} \simeq200$ km/s, and $t_g$ is the age of the galaxy $t_g\sim3\times10^{17}$s.

 A complication in estimating the number density of PBH is that they change with radial distance from the galactic centre $R$ (as do stars). However, if we assume that the PBH distribution follows an NFW profile, similar to dark matter,
 then we can evaluate $n_{\rm PBH}$ at a particular radius. 
 At $R=8$ kpc the local dark matter density is $\rho_{\rm DM}(R)\simeq 10^{-21}$ kg/m${}^3$ and it follows that
\beq
n_{\rm PBH}(8~{\rm kpc})
\sim 10^{-8} {\rm pc}^{-3} \left(\frac{f_{\rm PBH}}{10^{-7}}\right)  \left(\frac{0.1M_\odot}{M_{\rm PBH}}\right)~.
\eeq  
Thus the number of PBH encounters at 8~kpc over the age of the galaxy is
 \beq
\Gamma_\bullet(8~{\rm kpc}) t_g
\simeq   10^{-5}
\left(\frac{f_{\rm PBH}}{10^{-7}}\right) \left(\frac{0.1M_\odot}{M_{\rm PBH}}\right)~.
\eeq 
The probability of multiple PBH encounters $P_\bullet$, which we present in Table \ref{table:tab2} shows the cases in which PBH are relatively rare with $f_{\rm PBH}\sim10^{-7}$ (this value motivated by eq.~(\ref{eq5})) and the case in which PBH comprise 1\% of dark matter. Similar to the stellar case in Table \ref{table:tab1}, we expect this to be overly simplified, but to provide some useful intuition. Observe that for $f_{\rm PBH}\sim10^{-7}$ the probability of multiple encounters is negligible, while for $f_{\rm PBH}\sim0.01$ multiple flybys are explected to be relatively common.

  
\begin{table*}
\centering 
\vspace*{5mm}
\begin{tabular}{|c | c| c|} 
\hline
\# PBH Flybys ($N$) ~~~&~~~ $P_{\bullet N}$ for $f_{\rm PBH}=10^{-7}$ ~~~&~~~ $P_{\bullet N}$ for $f_{\rm PBH}=0.01$ \\ 
\hline 
1 & $10^{-5}$   &  0.34\\
2 & $10^{-10}$  & 0.11\\
3 & $10^{-15}$  &  0.03\\
\hline 
\end{tabular}
\caption{\normalfont{Probability of multiple 0.1$M_\odot$ PBH close encounters for a given star, assuming these events are Poisson distributed.}\label{table:tab2} }
\end{table*}

\renewcommand{\thefigure}{A\arabic{figure}}



 \vspace*{13mm}

\end{document}